# A STUDY ON URBAN MOBILITY IN NORTHERN CITIES OF BANGLADESH

**Riffat ISLAM**[1] **and Md. Kamruzzaman**[2]

[1] Department of Civil Engineering, Rajshahi University of Engineering and Technology
Rajshahi, Bangladesh
Email: [1]riffat_sarzia@yahoo.com and [2]kzaman@ruet.ac.bd

**Abstract.** *The percentage of urban areas dedicated to streets and public spaces is a crucial feature of spatial planning of cities. By 2050 urban mobility will be one of the biggest confronts of cities around the globe. This study has been carried out to assess the ratio of public space allocated to the streets in some cities of Bangladesh. The length, width, area and number of street crossings have been counted for city core and its suburban area, as an indicator of the form and pattern of the street layout. In this exercise, the rest of the public space, like gardens and general public spaces for amenities, including sports, are not taken into account. The methodology of data collection has been through Google Earth and GIS software. For precise results, other more sophisticated software is essential. Nevertheless, even at these levels of precision, very interesting city patterns emerge. The findings of this study show that Bogra, Rangpur, Dinajpur are better cities having higher land allocation rate for street, including sufficient crossings. Nilphamari and Thakurgaon have lower land area for streets; tend to have lower connectivity and productivity. Rest of the cities has average land area for street and average number of street crossings.*

## 1 INTRODUCTION

Urban mobility often used to indicate a set of interrelated measures to satisfy the mobility needs of people and businesses, today and tomorrow; in city and their surrounding areas. Urban mobility can ensure accessibility to the transport system to all, improve safety and security, reduce pollution and can make the urban environment attractive. It is one of the toughest challenges that cities face today as existing mobility systems are close to breakdown (Little A., 2014). The question of how to enhance urban mobility while at the same time reducing congestion, accidents and pollution is a common challenge to all major cities in the world.

National guideline for urban mobility planning provides orientation to local authorities. In several countries, such as Brazil, France and India, the development of urban mobility plans has become an obligatory requirement for receiving national government funds for local transport projects (GIZ, 2014). Too often, transport infrastructure fails to keep up with the mobility needs of the growing urban population. Setting a city on a sustainable course regarding its land use and transport system requires a clear roadmap of urban mobility plan. A successful urban mobility plan can provide a feasible and powerful strategy to tackle urban mobility challenges.

The percentage of urban areas dedicated to streets and public spaces is a crucial feature of urban mobility. Cities that have adequate street and public spaces and greater connectivity are more livable and productive (UN-Habitat, 2013a). The significance of urban mobility plan in providing solution for urban transport system in Bangladesh holds in stark contrast with the very limited research on the topic. Lack of information on characteristics and scale of 'Urban Mobility' makes it difficult to formulate appropriate policies and this paper attempts to uncover these facts.

## 2 OBJECTIVE OF THE STUDY

Mobility flows have become a key concern of urbanization. Despite the increasing level of urban mobility worldwide, access to places, activities and services has become increasingly difficult (UN-Habitat, 2013b). Many of the world's cities face an unprecedented accessibility crisis, and are characterized by unsustainable mobility systems.

The aim of the study is to examine the state of urban mobility in some selected cities of Bangladesh. It attempts to explore the linkages between percentage of urban land allocated for street network and level of urban mobility. The study also intends to highlight the status of different mobility parameters of each cities and their interrelation on the mobility ranking of all cities. Finally, the study plans to compare the state of urban mobility among Bangladeshi cities with some international cities.

## 3 METHODOLOGY

Percentage land area for street, street density and number of intersection are the parameters under which mobility ranking of study cities are conducted to propose whether a city is highly, moderately or poorly accessible. The study used orthophotos and GIS in determination of public space allocated to street in ten northern cities of Bangladesh. The number of street crossings also counted during the study. Images from 'Google Earth' were analyzed in AutoCAD first to find out the street length, street area and number of crossings.

Dinajpur, Bogra, Lalmonirhat, Pabna, Sirajganj, Rangpur, Nawabganj, Rajshahi, Nilphamri, Thakurgaon city are selected as study area. At the beginning, orthophotos of those cities are captured from Google Earth. Shape file of a particular city (under study) is collected and used in GIS software (Arc View) to convert it into KML file for projection on Google Earth. While considering a city under study, city core and suburb areas are considered for mobility analysis. City core is considered taking a particular radius around the buildup city centre. A greater study area has been considered ranging from city center to suburb to compare on the mentioned parameters. Using "Add Path" tool in Google Earth, street layout of all visible streets is drawn from Google Earth. The entire street network of study area including city core and suburb was exported to Auto CAD by converting the KML file to DXF file to work in AutoCAD platform. The map was drawn in proper scale. The entire area (city core and suburb) is divided into number of small blocks, using separate 'hatches', in Auto CAD. Street length is determined using '*tlen*' AutoLISP program. Land area and street area are computed for individual small blocks and the sum of all block determines total urban land area for street and the percentage of land allocated to street.

## 4 FIXATION OF STREETS IN CITY CORE AND SUBURBAN AREA

The city boundary is fixed by exporting the shape file of the corresponding city in Google Earth. For example, fixing of city core including suburb for Dinajpur city is shown in Fig. 1, 2 & 3 respectively.

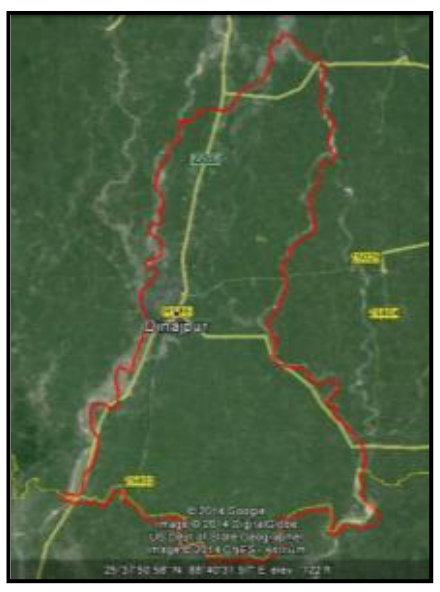

Fig 1: Ortho Photo of Dinajpur from Google Earth

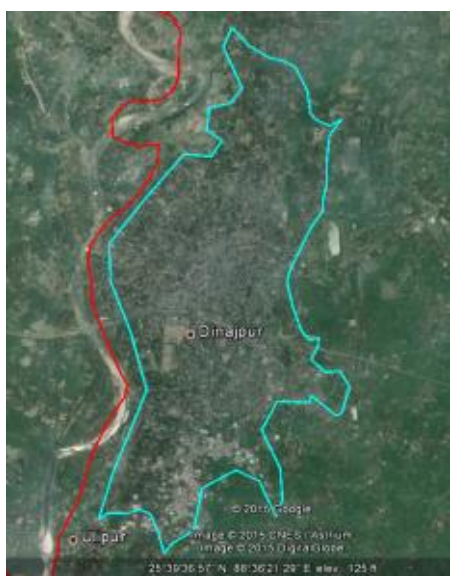

Fig 2: Suburb in Blue Color

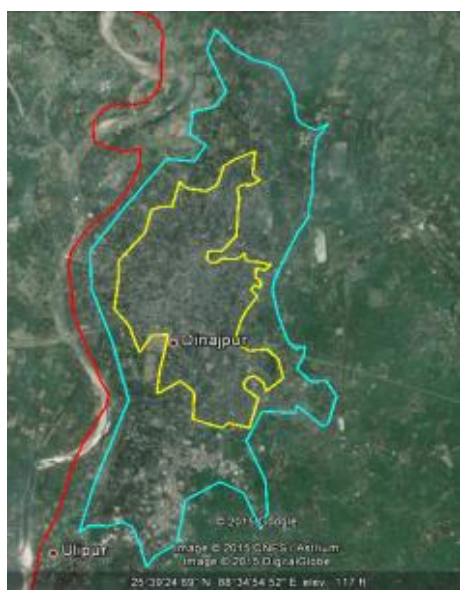

Fig 3: City Core in Yellow Color

Table 1 shows fixation of city core and suburb for Dinajpur city as an example:

Table 1: Fixation of City Core and Suburb Area (Dinajpur)

| City Name | City core area (km²) | Suburb area (km²) | Remarks | |
|---|---|---|---|---|
| Dinajpur | 5.35 | 15.97 | City core is considered taking 1.0 km radius in 'X' direction & 2.0 km radius in 'Y' direction from the city center. | Suburb is considered taking 1.35 km radius in 'X' direction & 4.00 km in 'Y' direction from the city center. |

## 5 COMPUTATION OF STREET LENGTH AND STREET AREA

Entire street layout is drawn in Google Earth (Fig 4). Length, width and land area of all streets and number of crossings are counted after the map is exported to AutoCAD (Fig 5) from Google Earth. Fig 6 shows street layout, hatched areas and street-intersections respectively for Dinajpur city.

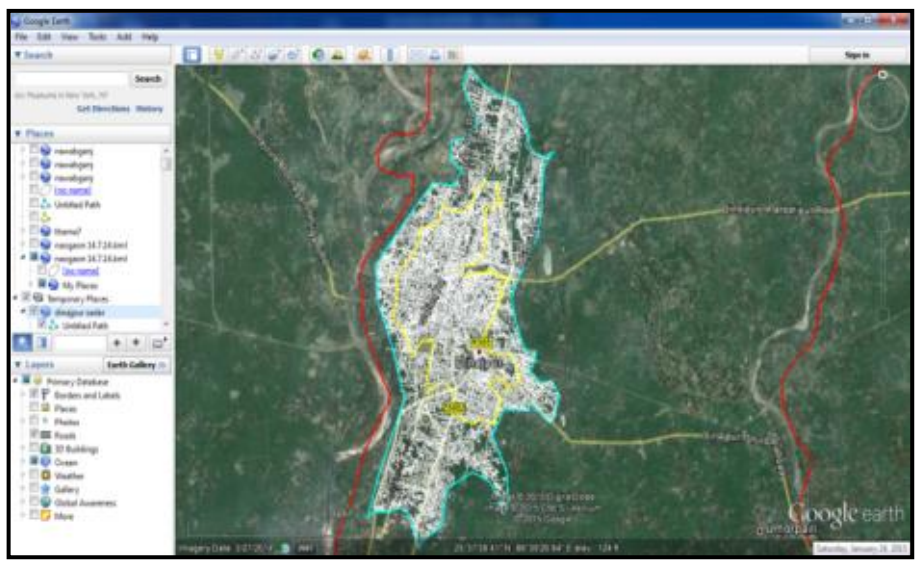

Fig 4: Entire Street Network of Dinajpur City is drawn in Google Earth.

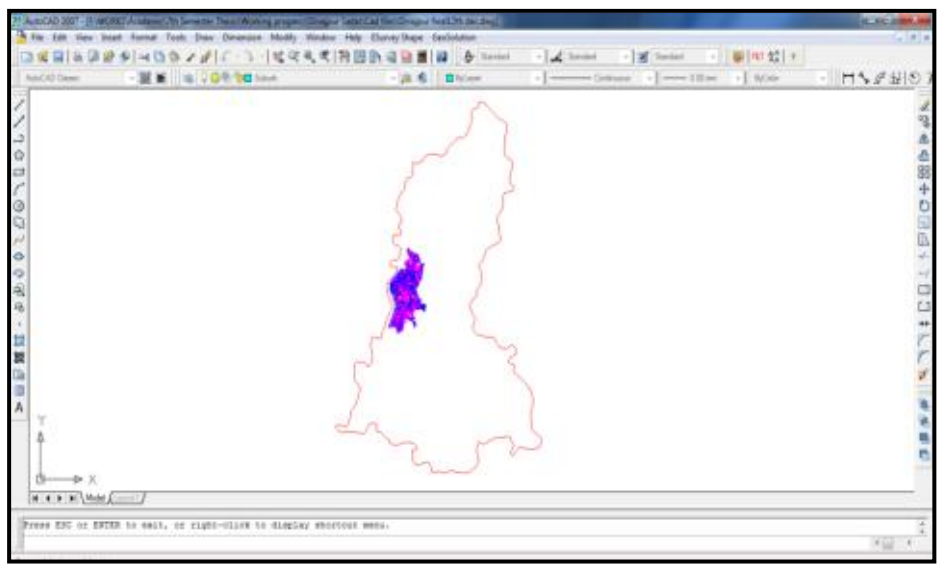

Fig 5: Complete Street Network of Dinajpur City is exported to AutoCAD for Analysis

Following the steps, street databases of all cities are prepared. For example street database for Dinajpur is presented in Table 2.

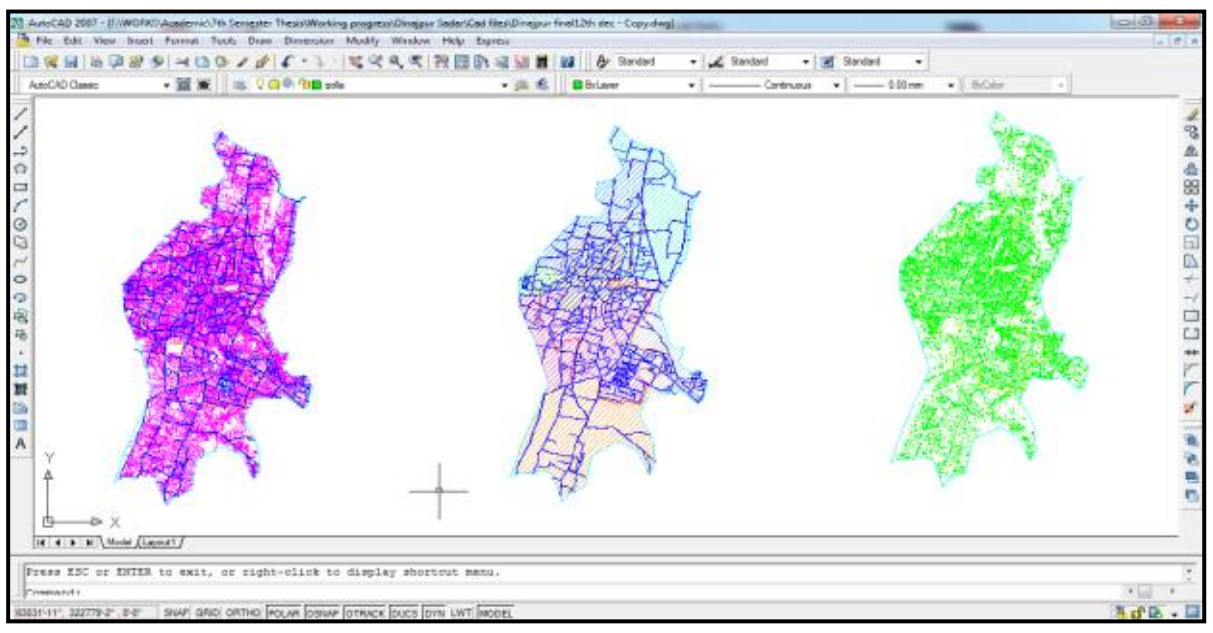

Fig 6: Street Network, Blocks and Marking of Crossings for Dinajpur City using Auto CAD.

Table 2: Percentage of Land Allocation for Street & Street Density for Dinajpur

| Region | Total Land Area (km²) | Total Street Area (km²) | Street Width (m) | Percentage of Street Area (%) | Total Streets Length (km) | Street Density (km/km²) | Total No. of Crossing | Intersection Density (Crossing/km²) |
|---|---|---|---|---|---|---|---|---|
| City core | 4.07 | 0.34 | 3.64 | 8.35 | 93.45 | 22.96 | 767 | 188.45 |
| City core & Suburb | 15.81 | 0.93 | 3.26 | 5.88 | 285.41 | 18.05 | 1964 | 124.23 |

## 6 RESULTS & DISCUSSION

### 6.1 Mobility Gradation of Cities

Table 3 shows a very important disparity of land allocation rate for street network, varying from 8.91 to 18.75 percent. This disparity itself is a good entry point to describe the urban mobility consequences in cities in Bangladesh. The top five cities ranges between 16.35 and 18.75 percent of urban land allocated to the street and may be considered as cities having good mobility. Land allocation rate falls between 14.18 and 15.94 percent of some cities which are considered having average mobility. Finally, cities having land allocation rate between 8.91 and 9.03 percent, may be considered as cities having poor mobility. Table 4 shows above average, average and below average ranking of cities based on the criteria of street density. Fig 7 and Fig 8 are graphical representation of ranking of cities based on percentage of urban land for street and street density. In both cases, Dinajpur and Bogra city are top ranking cities having street density and percentages of street area holds above average range. On the other hand, Nilphamari and Thakurgaon city has the lowest ranking having street density and percentages of street area having below average range. The remaining cities like Rajshahi (Boalia Thana), Sirajganj, Rangpur, Pabna, Nawabganj and Lalmonirhat has average accessibility and mobility.

Table 3: Ranking of Cities Based on Percentage of Urban Land for Street

| Name of Cities | Area of City Core (km²) | Total Street Area (km²) | Percentage of Street Area (%) | Ranking Based on % of Street Area | Remarks |
|---|---|---|---|---|---|
| Bogra | 4.16 | 0.78 | 18.75 | 1 | Above Average |
| Rangpur | 4.52 | 0.85 | 18.72 | 2 | |
| Dinajpur | 5.35 | 1.00 | 18.69 | 3 | |
| Sirajganj | 3.47 | 0.58 | 16.71 | 4 | |
| Pabna | 3.12 | 0.51 | 16.35 | 5 | Average |
| Rajshahi (Boalia) | 3.2 | 0.51 | 15.94 | 6 | |
| Lalmonirhat | 2.89 | 0.44 | 15.22 | 7 | |
| Nawabganj | 5.5 | 0.78 | 14.18 | 8 | |
| Nilphamari | 5.76 | 0.52 | 9.03 | 9 | Below Average |
| Thakurgaon | 18.4 | 1.64 | 8.91 | 10 | |

Table 4: Ranking of Cities Based on Street Density

| Name of Cities | Area of City Core (km²) | Length of Street (km) | Street Density (km/km²) | Ranking Based on Street Density | Remarks |
|---|---|---|---|---|---|
| Dinajpur | 5.35 | 309.4 | 57.83 | 1 | Above Average |
| Bogra | 4.16 | 238.58 | 57.35 | 2 | |
| Lalmonirhat | 2.89 | 146.38 | 50.65 | 3 | |
| Pabna | 3.12 | 153.84 | 49.31 | 4 | Average |
| Sirajganj | 3.47 | 164.15 | 47.31 | 5 | |
| Rangpur | 4.52 | 203.08 | 44.93 | 6 | |
| Nawabganj | 5.50 | 240.61 | 43.75 | 7 | |
| Rajshahi (Boalia) | 3.20 | 117.49 | 36.72 | 8 | |
| Nilphamari | 5.76 | 160.27 | 27.82 | 9 | Below Average |
| Thakurgaon | 18.4 | 406.02 | 22.07 | 10 | |

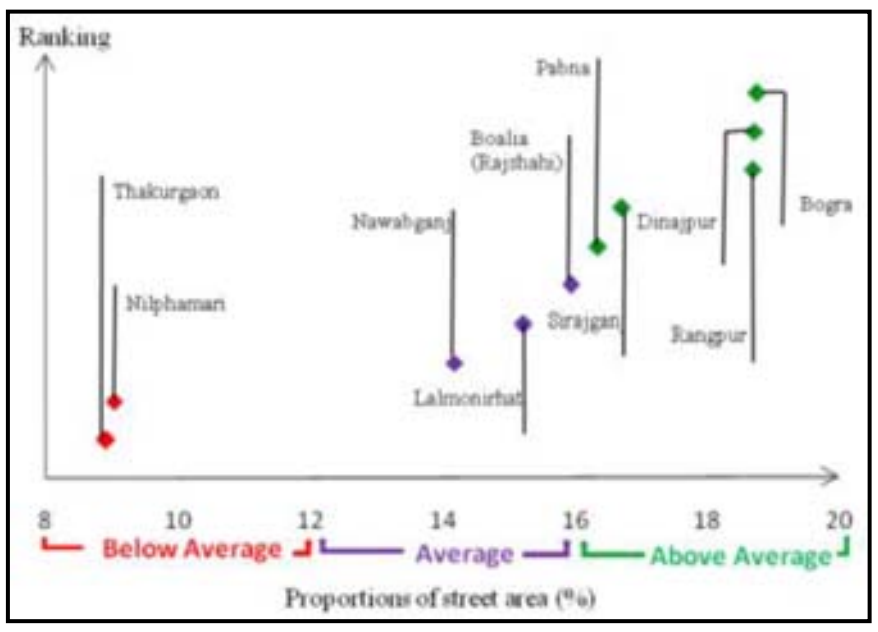

Fig 7: Ranking of Cities in Based on Percentage of Land for

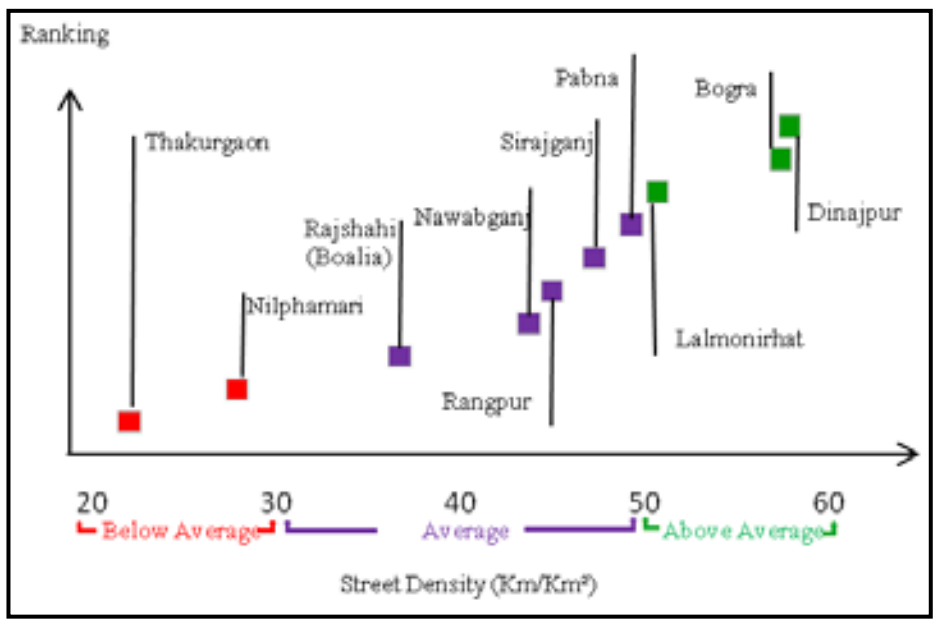

Fig 8: Ranking of Cities Based on Street Density

## 6.2 Relation between Percentages of Street Area and Intersection Density

Figure 9 establishes the inter-connection of the two variables studied above. The ranking of cities is provided for each variable and the linkage between the two variables. Ten cities have been chosen in order to have a wide variety of cities.

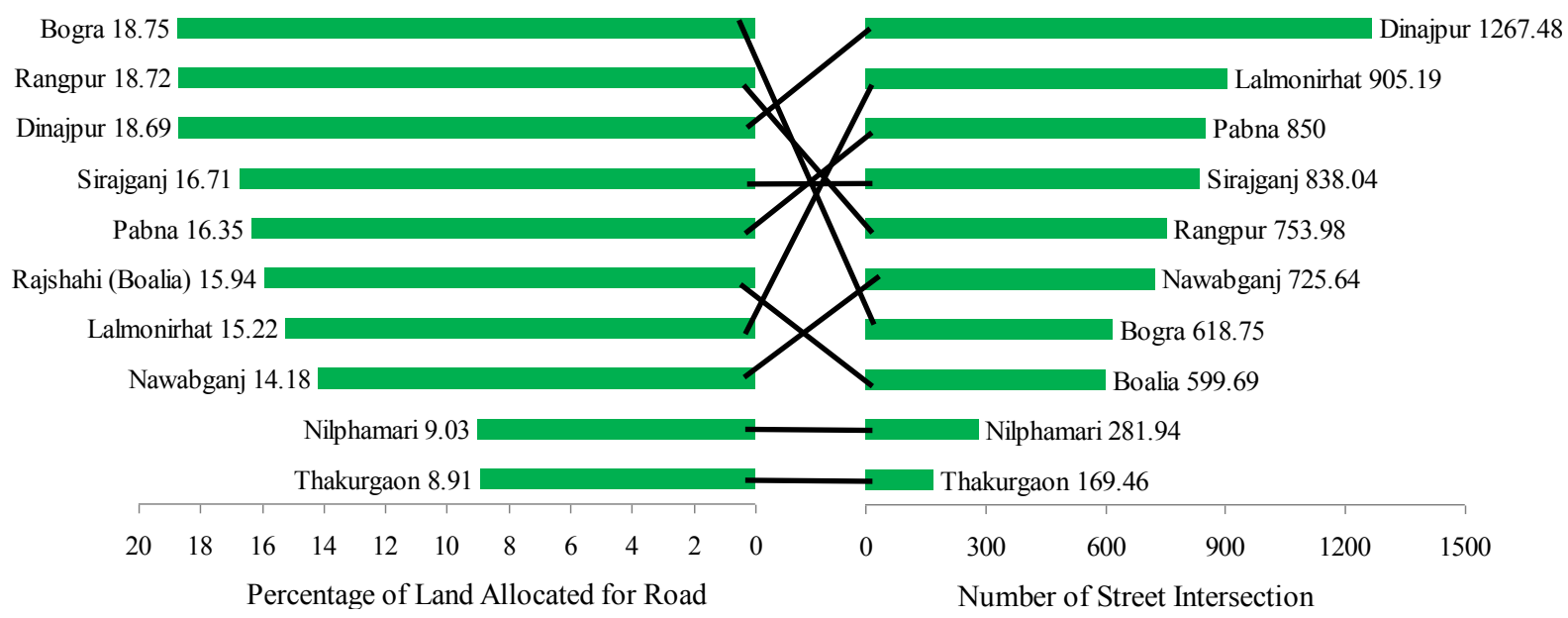

Fig 9: Inter-relation between Percentages of Street Area & Intersection Density

The central area of the city has been chosen, meaning areas of less than 5.0 km$^2$, in order to ensure the comparability of different cities. This provides for an interesting mixture of space and pattern which explains about the efficiency of the city. The result explores these factors in order to develop efficient tools of plotting for the matrix of the public space in the urban areas.

## 6.3 Frequency of Cities based on the Mobility Parameters

It is important to understand that the quantity of the public space is not related to the price of land. In general, the price of land is much higher in cities that have more public space. This should be an additional reason to ensure proper delinea-tion of unbuilt space at the beginning of urbanization, when the price of land is cheaper. Figure 10 shows that there are about 5 cities having land area more 16% for street network and only 2 cities having lass than 12% land for street. Cities are more accessible and attractive if the street area and open space are more.

Figure 11 displays the number of street intersections per km$^2$ of urban area in selected cities as measured by the described methodology. Cities range from 1267 (Dinajpur) to 169 (Takurgaon) crossings per km$^2$. Cities having more crossing are considered walkable and appropriate in many cities, in order to generate street life and for moving goods and services productively and efficiently. Majority of cities of Bangladesh have good number of intersection.

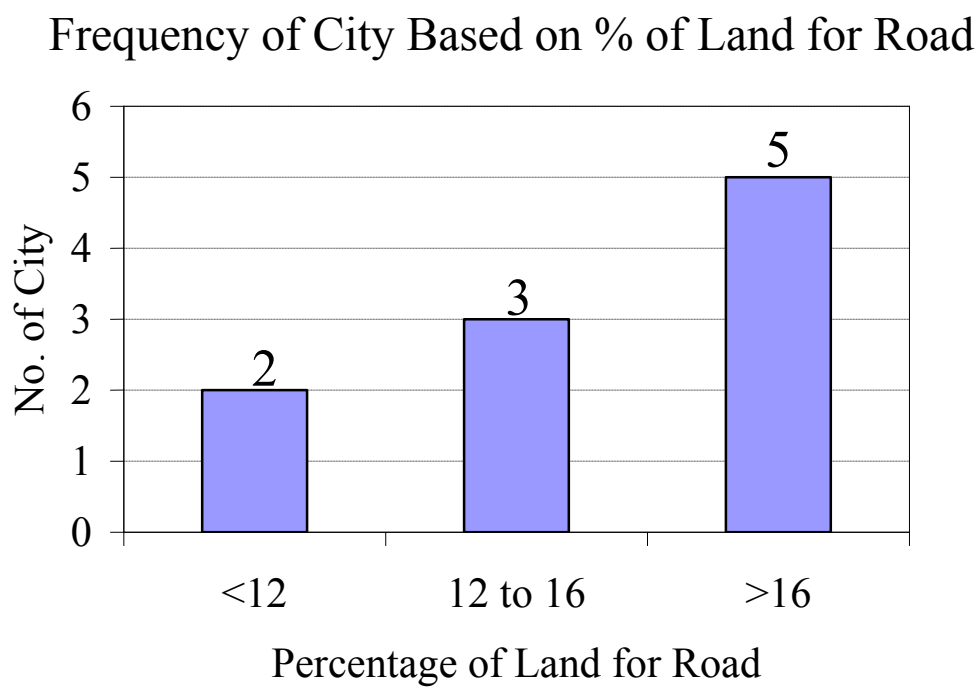

Fig 10: Frequency of Selected Cities having Different Percentage of Land for Street

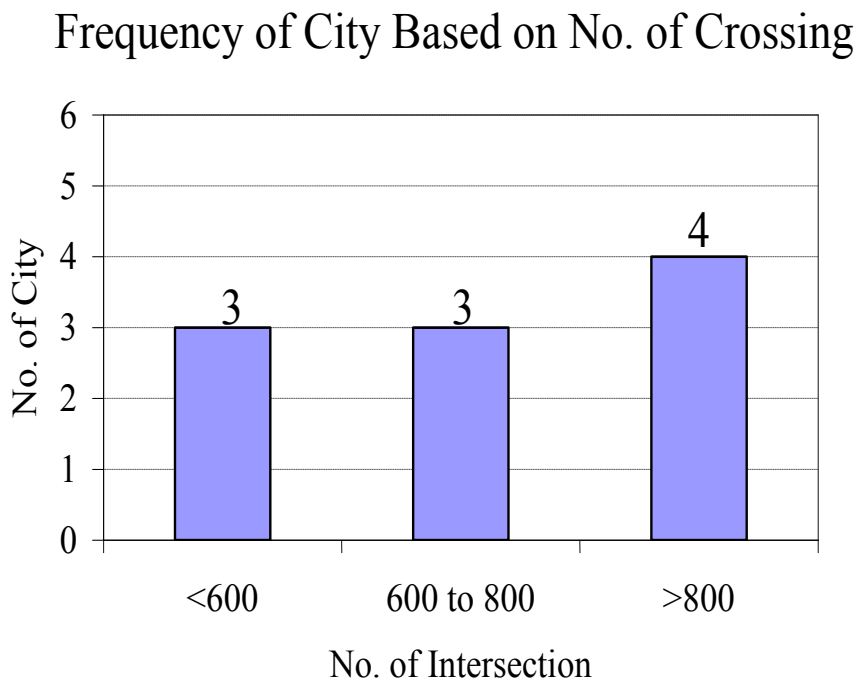

Fig 11: Frequency of Selected Cities having Different No. of Intersections

### 6.4 Comparison between International and Cities of Bangladesh

Fig 12 provides a clear representation of scale of urban mobility among various international cities and Northern cities of Bangladesh based on percentages of land area for street and number of crossing respectively. In general, international cities have more urban lands allocated for street and more intersections, which causes better mobility.

## 7 CONCLUSION

Ranking of cities according to the parameters gives a clear depiction of how mobility strengthens the accessibility of a high ranked city (Bogra) whereas weakens that of a low ranked city (Thakurgaon). These findings provides an entry point for addressing the issue of urban mobility, inefficiency or lack of urban planning and clearly suggest the high priority that cities in Bangladesh should put on early attention to planning when land is still inexpensive, in order to avoid future gridlocks and congestion.

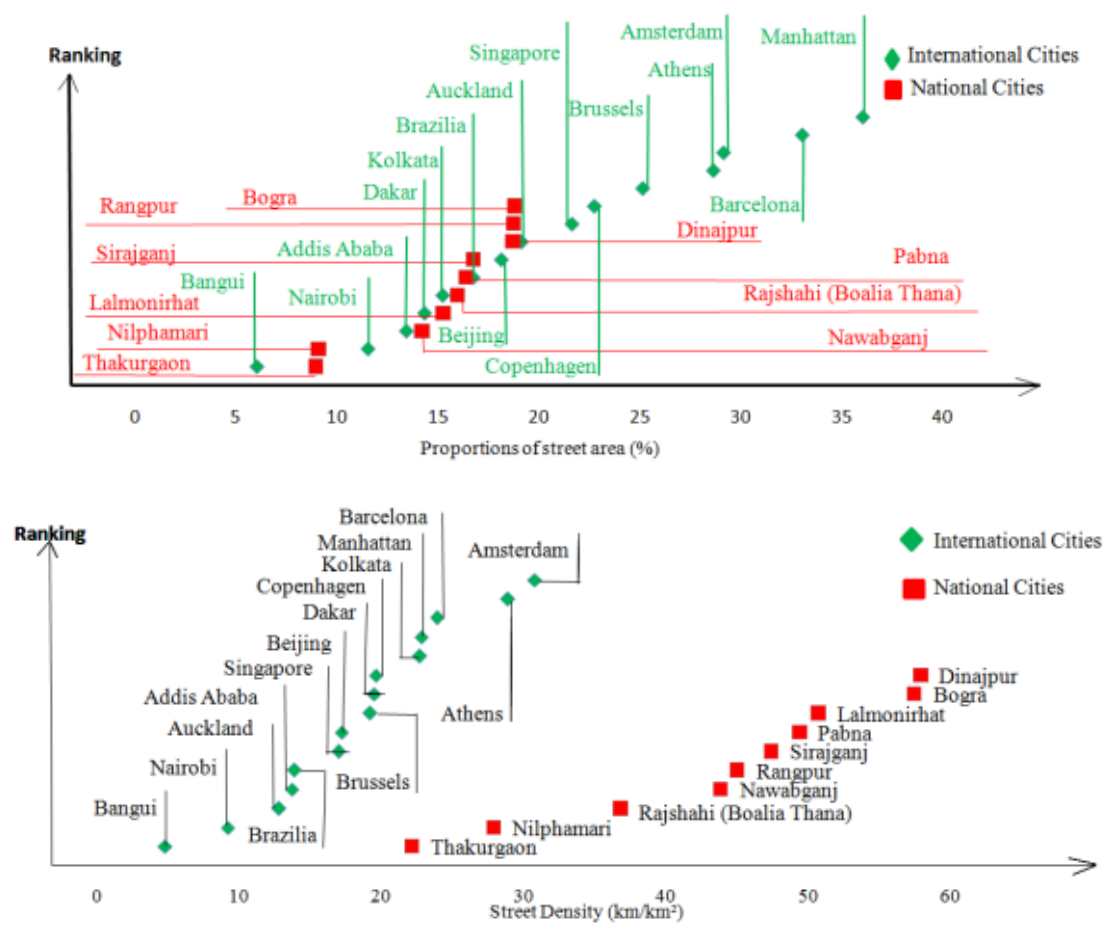

Fig 12: Ranking of 15 Different Cities of World and 10 Cities of Bangladesh in Relation to (Top) Percentage of Land Area for Street and (Bottom) Street Density.